\documentstyle[pra,aps]{revtex}

\newcommand{\be}{\begin{equation}}
\newcommand{\ee}{\end{equation}}
\newcommand{\lb}{\label}
\newcommand{\en}{\varepsilon}
\newcommand{\bk}{{\bf k}}
\newcommand{\bv}{{\bf v}}
\newcommand{\br}{{\bf r}}
\newcommand{\bff}{{\bf f}}
\newcommand{\bz}{{\bf 0}}
\newcommand{\bZ}{{\bf Z}}
\newcommand{\bE}{\hat{\bf e}}

\begin{document}
\draft


\title{\bf Analogies Between Scaling in Turbulence, Field Theory
and Critical Phenomena}
\author{Gregory Eyink and Nigel Goldenfeld}
\address{Physics Department and Beckman
Institute, University of Illinois at Urbana-Champaign,
1110 West Green Street, Urbana, Il. 61801-3080}

\maketitle
\begin{abstract}
We discuss two distinct analogies between turbulence and field theory.
In one analogue, the field theory has an infrared attractive
renormalization-group fixed point and corresponds to critical
phenomena.  In the other analogue, the field theory has an ultraviolet
attractive fixed point, as in quantum chromodynamics.
\end{abstract}
\vspace{0.3in}



In this brief note, we shall discuss two distinct analogies between
turbulence at high Reynolds number, statistical field theory and
critical phenomena.  Such analogies are implicit in attempts to
make a useful connection between these apparently disparate phenomena,
but a number of points deserve to be made explicitly.  Some of our
remarks are essentially trivial observations; nevertheless their
ramifications have not always been respected by approximate
theories \cite{0}.

The fully-developed turbulent regime is specified by the condition that
the integral length-scale $L$ of the largest eddies should be much
larger than the length scale $\eta_d$ at which dissipation is
effective.  The latter is usually defined as the inverse of the
wavenumber $k_d$ at which occurs the peak of the enstrophy spectrum
$k^2E(k).$ Equivalently, the Reynolds number ${\rm Re}\equiv UL/\nu$
should be large, where $U$ is a characteristic large-scale velocity and
$\nu$ is the molecular viscosity.  We consider the statistical
steady-state with constant mean rate $\bar{\en}$ of energy injection by
the turbulence production mechanism, which, by stationarity, is also
the
mean rate of energy dissipation by molecular viscosity.  We shall take
the point of view that the problem of turbulence is to understand the
properties of the stationary (but presumably non-Gibbsian) probability
distribution governing the velocity field of a turbulent fluid; it is
this distribution which is used below when we write averages in the
context of turbulence.

The first analogue of turbulence is to field theories with an infra-red
attractive renormal\-ization-group (RG) fixed point. This is the
situation
for field theories corresponding to critical lattice spin-systems. For
example, consider a system of spins $\sigma(\br)$ in $d$ spatial
dimensions, where $\br$ runs over a lattice $a\bZ^d$ of
lattice-constant
$a,$ governed by a (dimensionless) Hamiltonian of the form:
\be H[\sigma]=\sum_{\br\in a\bZ^d}a^d\,
\left[{{K}\over{2}}\sum_{\mu=1}^d
\left({{\sigma(\br+a\bE_\mu)-\sigma(\br)}
\over{a}}\right)^2+{{\tau}\over{2}}\sigma^2(\br)+{{\gamma}\over{4}}\sigma^4(\br)
+\cdots\right],
\lb{1} \ee
with $\bE_\mu$, $\mu=1,2, \dots, d$ being unit lattice vectors. The
parameter $\tau$ depends upon the temperature of the spin system, with
an order-disorder transition occuring at the critical value $\tau_c.$
In
the critical regime ${{(\tau-\tau_c)}/{\tau_c}}\ll 1$ there is, for
large space-separation, universality from the specific {\em
short-distance}, or lattice-scale interactions.
On the other hand, turbulence scaling in the high
wavenumber inertial-range is believed to exhibit universality from the
{\em small wavenumber} stirring mechanisms. Therefore, in this analogy
the roles of space and wavenumber are interchanged. A detailed list of
correspondences (and definitions) may be drawn up for an analogy along
these lines, which is motivated below:

\vskip 0.2 truein
\begin{tabular}{ll}

{\bf Turbulence} & {\bf Critical Phenomena} \\
space separation $r$ & wavenumber $k$ \\ viscosity, $\nu$ & temperature
variable
$\tau-\tau_c$ \\ energetic length-scale,
$L$ & UV cut-off $\Lambda,$ (or, inverse lattice-spacing, $a^{-1}$) \\
mean dissipation, $\bar{\en}$ & stiffness constant, $K$ \\ dissipation
wavenumber, $k_d\equiv \eta_d^{-1} $ & correlation length $\xi$
\\ velocity correlation
function, & spin correlation function, \\
$S_2(\br)=\langle(\bv(\br'+\br)-\bv(\br'))^2\rangle$ &
$C(\bk)=\sum_{\br\in
a\bZ^d}a^de^{i\bk\cdot\br}\langle\sigma(\br)\sigma(\bz)\rangle$ \\
intermittency exponent, $\mu$ & correlation exponent, $\eta_c .$ \\

\end{tabular}

\vskip 0.2 truein
\noindent
This analogue is similar to that pointed out previously by Nelkin
\cite{1}, deGennes \cite{2} and Rose and Sulem\cite{3}, the critical
($\tau\rightarrow\tau_c$) limit of equilibrium spin systems being
considered analogous to the zero-viscosity limit of turbulence.
However,
we have made a correspondence of the constants $\bar{\en}$ and $K$
which
does not seem to have been pointed out before. Also, we propose that
$\mu$ and $\eta_c$ are directly analogous, rather than being related as
$\mu\leftrightarrow {{2}\over{3}}-\eta_c,$ as suggested by the earlier
authors.

The above analogy can be motivated by comparing Kolmogorov's 1941
theory of
turbulence \cite{4} and Landau's 1937 ``mean-field'' theory of critical
phenomena\cite{5}.  In both theories, critical exponents are obtained
by
dimensional analysis \cite{6}.

In the Landau mean-field analysis, it is implicitly assumed that the
long wavelength properties are independent of the lattice constant
$a$.
Thus the limit $a\rightarrow 0$ of spin-correlations is presumed to
exist, an assumption that the asymptotics are of the first kind
\cite{barenblatt}. In that case the only remaining parameter with units
of length is $r_0^{-1/2}=\left({{(\tau-\tau_c)}/{K}}\right)^{-1/2},$
yielding the Landau prediction for the correlation length $\xi_L\sim
r_0^{-1/2}.$ The values for the critical exponents
follow from this assumption: for example,
\be C(\bk)\sim K\cdot k^{-2}, \lb{2} \ee
for $k\xi\gg 1$, where it has also been assumed that the critical
limit $r_0\rightarrow 0$  exists.  In fact, Landau's first
assumption that the limit $a\rightarrow 0$ exists is generally wrong,
and the correct asymptotic scaling behavior is of the second
kind\cite{barenblatt} \be C(\bk)\sim K\cdot k^{-2}(ka)^{\eta_c}, \lb{3}
\ee for some $\eta_c>0,$ in which the microscopic length $a$ appears
explicitly. Likewise, the Landau scaling of the correlation length has
an ``anomalous'' correction arising from $a$-dependence:
$\xi\sim\xi_L\cdot(r_0 a^2)^\theta$ for some $\theta>0,$ or $\xi\sim
r_0^{-\nu_c}$ with $\nu_c={{1}\over{2}}-\theta.$

In the same way, Kolmogorov assumed that the limit $L\rightarrow
+\infty$ should exist in turbulence with finite velocity correlations.
In that case, the only remaining length-scale is the dissipative scale
estimated as $\eta_K\sim (\bar{\en})^{-1/4}\nu^{3/4}.$ Thus, Kolmogorov
obtained by dimensional analysis
\be S_2(r)\sim (\bar{\en}r)^{2/3}, \lb{4} \ee
when $k_d r\gg 1,$ if the limit $\nu\rightarrow 0$ exists
for velocity correlations.  As with the corresponding assumption in
Landau's theory, Kolmogorov's hypothesis of existence of the first
$L\rightarrow +\infty$ limit is questionable, due to the build-up of
fluctuations of energy flux in the energy cascade. (Ironically, this
criticism originates in part from a remark of Landau himself
\cite{7}!).
Instead, simple cascade models indicate that the scaling law should
instead be of the form \be S_2(r)\sim
(\bar{\en}r)^{2/3}\left({{L}\over{r}}\right)^{-\mu}. \lb{5} \ee for
some
$\mu>0.$ Similarly, these models suggest \cite{7x} that the dissipation
scale may have a slight dependence upon $L,$ as $\eta_d\sim
\eta_K\left({{L}\over{\eta_K}}\right)^\delta$ for $\delta\neq 0,$
leading to a scaling $\eta_d\sim\eta_K^\omega$ with $\omega=1-\delta$.

It is clear that in these formulas, $K, a,\eta_c,\nu_c$ are homologous,
respectively, to $\bar{\en}, L^{-1},\mu,\omega.$ Notice that typically
$\eta_c$ is small in critical systems even in two or three dimensions
and that $\mu$ represents an empirically small modification to the
dimensional analysis result of 5/3 for the exponent appearing in the
Fourier transform of $S_2$:  \be E(k)\sim
(\bar{\en})^{2/3}k^{-5/3}(kL)^{-\mu}. \lb{6} \ee The constants $K$ and
$\bar{\en}$ play a similar formal role in the two theories.
To see this, consider a
Martin-Siggia-Rose (MSR) Lagrangian \cite{7a,7b} for the steady-state
turbulent cascade state produced by driving with a Gaussian random
force
$\bff,$ white noise in time, with zero mean and covariance \be \langle
f_i(\br,t)f_j(\br',t')\rangle= 2\delta_{ij}F(\br-\br')\delta(t-t').
\lb{7} \ee It is easy to check that $F$ has the units of energy
dissipation and, indeed, it is directly related to the mean energy
injection rate by $\bar{\en}=2F(0)$.  Therefore, the quadratic term in
the MSR action corresponding to the ``response field'' $\hat{\bv}$ is
\be S^{(0)}[\hat{\bv}]={1\over 2} \int dt\,\int
d\br\,d\br'\hat{\bv}(\br,t) F(\br - \br') \hat{\bv}(\br',t) , \lb{9}
\ee
which is proportional to $\bar{\en}$. $S^{(0)}$ has a formally similar
structure to the quadratic term in the spin-Hamiltonian, which is
proportional to $K.$

A second, rather different analogy can be made between turbulence and
field theory with an ultraviolet (UV) attractive fixed point, or, field
theory for short. (It is generally accepted that, in order for a
continuum field theory model to be well-defined, it must be
``asymptotically safe,'' i.e. it must correspond to a lattice model
with
a UV attracting fixed point.)  Field theory, such as UV asymptotically
free quantum chromodynamics (QCD), exhibits scaling at short
distances, just as turbulence is believed to do. Therefore, in this
analogy space corresponds to space. The spin system we examined before
may still be used in this analogy, when it is considered in dimension
$d < 4$ along the single RG trajectory which flows in the UV direction
into the non-Gaussian Wilson-Fisher fixed point (previously
we considered the theory at a generic point on or slightly away from
the
critical surface). It is convenient to use
the field-theoretic notation, introducing a (lattice) field \be
\phi(\br)=K^{1/2}\sigma(\br), \lb{10} \ee which has dimension
${{1}\over{2}}(d-2)$ in units of inverse-length. As before, a detailed
list of correspondences may be drawn up:
\vskip 0.2 truein

\begin{tabular}{ll}
{\bf Turbulence} & {\bf Field Theory} \\
space separation $r$ & space separation $r$ \\
viscosity $\nu$ (or, Kolmogorov scale, $\eta_K$) & lattice-spacing $a$
\\
energetic length-scale, $L$ & correlation length $\xi$ \\
Kolmogorov wavenumber, $k_\eta$ & UV cut-off $\Lambda=a^{-1}$ \\
velocity correlation function, & field-theoretic Green function, \\
$S_2(\br)=\langle(\bv(\br'+\br)-\bv(\br'))^2\rangle$ &
$G(\br)=\langle\phi(\br)\phi(\bz)\rangle$ \\
\end{tabular}

\vskip 0.2 truein
\noindent
These correspondences essentially all follow from the comparison in the
last line.  For example, the length-scale $L$ gets its name
``integral-scale'' from the fact that it is an integral correlation
length of the longitudinal velocity correlation $f(r)=
\langle\left(\hat{\br}\cdot[\bv(\br'+\br)-\bv(\br')]\right)^2\rangle,$
defined by \be L={{1}\over{f(0)}}\int_0^{\infty}dr\,f(r), \lb{11} \ee
Therefore, it corresponds naturally to the correlation length $\xi$
defined by the exponential decay rate of $G(\br).$ Likewise, the
dissipation length $\eta_K$ provides a short-distance cut-off for
turbulent velocity correlations, analogous to the lattice cutoff $a$
for
the field theory correlations. The correspondences made here,
$(k_\eta,L)\leftrightarrow (\Lambda,\xi),$ are the opposite of those
made in the first analogy, $(k_\eta,L)\leftrightarrow (\xi,\Lambda).$
In
the present analogy, the zero-viscosity limit is analogous to the
continuum limit $a\rightarrow 0$ of field theory.

Having pointed out the analogies between turbulence and field theory,
it
is appropriate to emphasize certain important distinctions.  First, we
discuss the issue of universality.  In turbulence, short-distance
scaling is believed to be generic and essentially independent of
large-scale statistics or driving mechanisms. Scaling laws, strictly
speaking, require an inertial range of infinite extent, so that we
confine our discussion here to the idealized situation of zero
molecular
viscosity (i.e. the critical limit according to our first analogy). We
will assume that the stationary probability distribution governing the
turbulent state is a fixed point in a function space, whose axes may be
thought of as coupling constants for all conceivable local (and perhaps
non-local) operators. This distribution is usually assumed to be the
fixed point of a RG transformation which integrates out {\it short
wavelength\/} degrees of freedom, just as in critical phenomena.  (An
alternative to this procedure will be mentioned below.)  Short-distance
universality implies that the {\it ultra-violet\/} RG flows should all
be in towards the fixed point: this distribution is a global sink in
the
ultra-violet.  (Here again we ignore crossover phenomena associated
with
finite molecular viscosity or other short-distance regularization
mechanisms.) Thus, the usual infra-red RG flow diagram, moving the
system to longer and longer length scales, will contain a fixed point
with a presumably infinite number of relevant directions. The physical
significance of relevant directions, in the context of critical
phenomena, is that they represent the parameters that must be tuned in
order for the system to be at criticality (e.g. temperature is at the
critical temperature, external field is zero).  In the context of
turbulence, these relevant directions correspond to the myriad of
different large-scale stirring mechanisms which generate the same
short-distant behaviour.  One of these relevant directions,
corresponding to the temperature, is the ``eddy viscosity'', which is
the effective viscosity at a given scale generated by turbulent degrees
of freedom at shorter scales.  The eddy viscosity will tend to the
bare,
molecular viscosity (assumed zero) as the turbulence fixed-point is
approached at short length-scales. In contrast, even field-theories
with
a UV fixed point must generally be tuned to lie on the low-dimensional
surface which attracts to the fixed point, and as mentioned above, the
number of variables that must be tuned is equal to the number of IR
unstable directions at the fixed point: this is usually a finite
number.  This general argument does not distinguish between relevant
and
marginally relevant directions.

The second distinction is that scaling behavior may be quite different
in turbulence and typical field-theories. The latter usually show
simple
``gap-scaling '' or ``hyperscaling'' in which higher-order correlations
scale with exponents which are just integer multiples of a single field
dimension. For example, if $d_\phi$ is the scaling dimension of the
field $\phi$, then typically \be \langle
\phi(\lambda\br_1)\cdots\phi(\lambda\br_p)\rangle\sim \lambda^{-p\cdot
d_\phi},\quad p=1,2,3,\dots \lb{12} \ee as $\lambda\rightarrow 0.$ In
contrast, higher-order velocity structure functions, $S_p(r),$ are
believed to scale in high Reynolds number turbulence with $r\rightarrow
0$ as \be
\langle\left(\hat{\br}\cdot[\bv(\br'+\br)-\bv(\br')]\right)^p\rangle\sim
(\bar{\en}r)^{p/3}\left({{L}\over{r}}\right)^{-x_p}, \lb{13} \ee for
$x_p<0$ \cite{7c,7d}. This type of ``multifractal'' scaling is found in
very few field theories. In the cases where it occurs, it is usually
associated with sequences of field variables with {\em negative}
scaling
dimensions. More generally, the multifractal scaling occurs if there is
a sequence of variables ``additively coupled'' in the operator product
expansion and negative scaling dimensions are one common mechanism by
which such sequences are produced \cite{7e,7f}. A representative case
is
the $O(N)$ nonlinear $\sigma$-model in dimension $d=2+\en$ with $N<2$
\cite{8}, which has such a sequence. This example is closely related to
Wegner's theory of the mobility edge in disordered electron systems and
the hypothesis there of multifractal electron wavefunctions for the
localized states \cite{8}. Note that any variable with scaling
dimension
$x<d$ corresponds to an unstable direction at the fixed point.
Therefore, the variables with negative scaling dimensions correspond to
an infinite set of unstable directions, and the associated fixed point
is UV attractive in a large domain, as suggested in the preceding
paragraph.  Motivated by this type of example, an RG theory of high
Reynolds number scaling in turbulence can be developed \cite{9,10} in
which the ``anomalous dimensions'' are the negative scaling dimensions
$x_p$ of the powers of the {\em velocity-gradients}, $(\nabla\bv)^p$.
In
traditional turbulence terminology this corresponds to generalized
flatnesses for velocity gradients diverging as powers of the Reynolds
number, which is also observed in both simulation and experiment
\cite{7f,10a}. The same phenomena lead also to the possibility of a
hierarchy of dissipation lengths $\eta_d^{(p)}$ delimiting the
short-distance end of the scaling range of the $p$th structure
functions
$S_p,$ each having a different dependence on molecular viscosity
$\eta_d^{(p)}\sim \eta_K^{\omega_{(p)}}$ \cite{7x}.

The third distinction that we wish to mention is the form of the
continuum limit.  In field theory taking the continuum limit requires
making a multiplicative {renormalization} of the lattice
field-variables, \be [\phi](\br)=Z(a)\cdot \phi(\br), \lb{13a} \ee
where
$Z(a)\sim a^{-\gamma_\phi}$ and $\gamma_\phi$ is an ``anomalous
dimension'' of the renormalized field $[\phi].$ The necessity of this
renormalization is connected to the UV divergences in the
field-theoretic Greens functions of the ``bare'' fields $\phi$ (cf.
Eq.(\ref{3})). Unlike the fields $[\phi]$ which exist only as
distributions in the limit $a\rightarrow 0,$ the velocity field in
turbulence must remain an ordinary function in the limit
$\nu\rightarrow 0.$ The requirement of finite mean kinetic energy
$\langle {{1}\over{2}}v^2\rangle<\infty$ implies  this (since the
velocity field must then be locally square-integrable with probability
one).  On the other hand, the velocity gradients will exist necessarily
only as distributions and UV divergences may appear associated to their
products at a single space-point. Only these products require
``renormalization'' (by suitable powers of $\eta_K$) in the
zero-viscosity
limit and any ``anomalous dimensions'' in inertial-range turbulence
scaling must therefore be associated with the velocity gradients rather
than the velocities themselves.

The two analogies we have pointed out differ most obviously in the
reverse roles played by scale (wavenumber) and space. However, they
differ more essentially in terms of their motivation, the first (with
critical phenomena) being more suggestive in physical terms and the
second (with field theory) arising from mathematical considerations of
the formal renormalization procedure.

In our first analogy, the correlation length in critical phenomena and
the dissipation scale in turbulence are intrinsic properties of the
system, determined by its detailed dynamical and statistical
properties,
rather than fixed external inputs. In contrast, the lattice spacing in
the first case and the integral scale in the second case are parameters
fixed by the experimental setup or the definition of the model. The
inverse role of large and small scales in the two cases arises from the
different character of ``cascade'' in the two instances. According to
the cascade picture, there is a transfer of excitation on the average
from the large turbulent eddies to the small ones by a stepwise
process,
which is chaotic in nature and entails a loss of memory of the
large-scale statistics.  Wilson has emphasized \cite{10b} that there is
also a ``cascade of fluctuations'' in critical phenomena. Droplet
fluctuations nucleated at the lattice-scale in the critical state can
grow to the size of the correlation length (and vice versa). However,
it
is now the details of the lattice structure which are lost and the
scale-invariant distributions of the large ``droplets'' which are
universal.

These facts have suggested to several
authors\cite{11},\cite{12},\cite{13} that in constructing an RG theory
of turbulence, it may be more natural to reverse the usual procedure
and
to eliminate {\em low-wavenumber shells} rather than high-wavenumber
shells as in the Wilson method. Some cautionary remarks on such a
procedure at a physical level are made in \cite{14}. However, our
discussion of the field theoretic analogy shows also that to study
short-distance UV-scaling, an IR-elimination RG is {\em not} required.
Lattice QCD is a good case in point, where the {\em short-distance}
asymptotic freedom is studied by the same UV-elimination procedure as
used in critical phenomena. The UV-elimination RG is also closely
related to an important practical problem of turbulence theory, that of
``subgrid-scale eddy modelling,'' and is naturally described in terms
of
simple ``eddy viscosity'' concepts. Indeed, the UV-elimination RG is
deeply connected with such eddy viscosity ideas, and RG-invariance
under
this operation is nothing more than a restatement of the fact that the
inertial range theory is independent of the molecular viscosity
\cite{10}.

\bigskip
\noindent
{\bf Acknowledgements.} We would like to thank the participants of the
Aspen Workshop on Fully-Developed Turbulence in August 1993 for many
useful comments and discussions on the topic of this paper,
particularly
J. Eggers, M. Nelkin and P. B. Weichman.  We thank the last two for
suggesting that this work be written up.  We thank the Aspen Center for
Physics for its hospitality, and the National Science Foundation for
its
partial support of this work through grant number NSF-DMR-93-14938.

\end{document}